\documentclass[twocolumn,secnumarabic,amssymb, nobibnotes, aps, prl,reprint]{revtex4-1}
\usepackage{subfigure}
\usepackage{graphicx}
\usepackage{upgreek}
\usepackage{color}

\begin{document}

\title{Gas-solid phase transition in laser multiple filamentation}%

\author{D. Mongin}
\author{E. Schubert}
\author{N. Berti}
\author{J. Kasparian}
\author{J.-P. Wolf}
\email[corresponding author ]{jean-pierre.wolf@unige.ch}

\affiliation{Universit\'e de Gen\`eve, GAP, Chemin de Pinchat 22, CH-1211 Geneva 4, Switzerland}
\date{October 4, 2016}%

\begin{abstract}

While propagating in transparent media, near-infrared multi-terawatt (TW) laser beams break up in a multitude of filaments of typically 100-200 um diameter with peak intensities as high as 10 to 100~TW/cm$^{2}$. We observe a phase transition at incident beam intensities of 0.4~TW/cm$^2$, where the interaction between filaments induce solid-like 2-dimensional crystals with a 2.7 mm lattice constant, independent of the initial beam diameter. Below 0.4~TW/cm$^2$, we evidence a mixed phase state in which some filaments are closely packed in localized clusters, nucleated on inhomogeneities (seeds) in the transverse intensity profile of the beam, and other are sparse with almost no interaction with their neighbors, similar to a gas. This analogy with a thermodynamic gas-solid phase transition is confirmed by calculating the interaction Hamiltonian between neighboring filaments, which takes into account the effect of diffraction, Kerr self-focusing and plasma generation. The shape of the effective potential is close to a Morse potential with an equilibrium bond length close to the observed value. 

\end{abstract}

\maketitle


Many physical systems are well described by statistical models driven by nearest-neighbor interactions, such as spin models~\cite{Marro2005,Kosterlitz1973,Rogers1989}. Recently, we showed that the formation of multiple filamentation patterns in high-power ultrashort laser pulses also belong to this category~\cite{Ettoumi2015a,Ettoumi2015b}. 

Laser filaments are produced in high-power, ultrashort laser pulses propagating in air or other transparent media~\cite{ChinHLLTABKKS2005,CouaiM2007,BergeSNKW2007} by a dynamic balance between Kerr self-focusing, and defocusing by higher-order non-linear effects including the ionization of the medium and other polarization saturation terms \cite{BejotKHLVHFLW2010a,B'ejHLKWF2011a}. At 800 nm, this balance clamps the filaments intensity to typically 50 TW/cm$^2$~\cite{KaspaSC2000,BeckeAVOBC2001}.

Filaments cannot be considered as virtual optical fibers guiding the light within their core: they continuously interact and exchange energy with the photon bath surrounding them~\cite{LiuTAGBC2005,CourvBKSMYW2003,KolesM2004,SkupiBPL2004}. 
This photon bath, also known as \emph{energy reservoir}~\cite{LiuTAGBC2005,EisenPSZ2008,MillsKC2013}, plays a key role in the filamentation process. In particular, the photon bath feeds the filament and balances its energy losses, allowing it to self-heal after an obstacle~\cite{CourvBKSMYW2003,KolesM2004,SkupiBPL2004} or extend its propagation distance~\cite{scheller_externally_2014}.

In the case of multifilamentation, the laser energy reservoir also mediates interactions between neighboring filaments~\cite{Berge1997,RenHFDM2000,BergAGSW2003,HosseLFLCKPAK2004,MaLXGZ2008,D'AsaHPAGSA2009,StelmRPFKWW2009}. This interaction is attractive if the filaments are in phase, and repulsive if they are in antiphase~\cite{XiLZ2006,ShimSHVHIG2010}, corresponding to constructive and destructive interferences in the photon bath, respectively. Previous theoretical and experimental studies have shown that the relative phase between filaments is mainly randomly driven by intensity fluctuations during the collapse \cite{shim_loss_2012} and is then stabilized for the filaments propagating after the collapse~\cite{jhajj_spatiotemporal_2016}.

Recently we showed that at laser powers exceeding 100 TW, this mutual interaction limits the density of filaments in the transverse beam profile~\cite{HeninPKWJKBSSNSRWSMBS2010a}, resulting in the rise of the photon bath intensity. As a consequence, the photon bath effectively contributes to non-linear effects like white-light generation~\cite{PetitHNBJKBSSRKSWW2011} or laser-induced condensation~\cite{PetraHSBKSVVKSWWW2011}, resulting in an overall increase of the yield of these processes.

The unexpected limitation of the number of filaments in the beam cross-section by the available space instead of the available power induced us to investigate this phenomenon in more depth. Many previous studies (\cite{M'ejKYSFWSVNCB2005,SkupiBPLMYKSWRWBS2004,M'ecDATFPMCSS2005,B'ejBEMWALSSKBCGBBBCCHLMMMPPRR2007})reported indeed that the number of filaments is directly proportional to the available power, and more precisely that the formation of each filament would require typically 5 critical powers $P_{\textrm{cr}}$ ($P_{\textrm{cr}}$ is the minimum power ensuring that self-focusing overcomes diffraction~\cite{boyd_nonliear_1992}). Since the space-limited filament formation was observed at 100 TW, this suggested the existence of a threshold between 2 different regimes.

Here we show that this transition is generic (also for lower powers), that the driving parameter of this transition is intensity and not power, and that for NIR multi-Terawatt beams it occurs at an intensity of 0.4~TW/cm$^2$. Above this threshold, the density of filaments is not limited by the available power , but by their mutual interaction. In this situation, the local filament density saturates around 12~cm$^{-2}$, with fixed characteristic distances, which reminds the structure of a solid. The two regimes can thus be considered as two phases, one close to a gas, where filaments have negligible interaction, and one close to a solid, where distances are determined by the balance between attractive (Kerr effect) and repulsive interactions (diffraction and plasma generation). A transition between these two phases induced, in our case, by increasing the equivalent of a pressure, i.e. the light intensity over the beam section.
Our work therefore offers an analogy between atomic/solid state physics and multi-filamentation phenomena, where the organisation of filaments in the transverse beam of a given intensity is similar to that of molecules for a given pressure.



The evolution of the filament number in the beam section was investigated over 45~m in the free atmosphere with the Teramobile laser~\cite{WilleRKMYMSWW2002}, which produced at 800~nm either~130 fs pulses with an energy tunable from 10 to 100~mJ in an 8~mm diameter beam (FWHM) (0.15~–-~1.5~TW/cm$^2$), or 50 to 400~mJ, 180~fs pulses in a 30~mm beam (40~–-~275~MW/cm$^2$). Both beams were emitted collimated (i.e., unfocused).

Beam profiles of individual pulses were recorded by a digital camera (Nikon D90, Macro-Zeiss f=2/100~mm ZF.2 lens) through a long-pass 750~nm filter. Simultaneously, a pre-calibrated photodiode monitored the energy of each individual pulse in order to take into account the input energy fluctuations related to the laser operation (typically $\pm10\%$) and assign actual intensities to each recorded image. 

Data were recorded at four distances, upstream, close to and downstream of the non-linear focus, which was measured at 7~m and 20~m from the laser exit for the 8~mm and the 30~mm beam, respectively. Single-pulse beam profiles and pulse energies were recorded at each distance. Filaments were identified as bright disks of diameter between 0.2 and 0.5~mm [12 - 30 pixels], using a circular Hough transform (Matlab function \emph{imfindcircle}). The resulting filaments number and individual positions were used to compute the filaments surface density, as well as the nearest neighbor distance in each condition. These results were then averaged over 40 energy bins. The error bars displayed in the following figures show the standard deviation evaluated from these statistics. 
To get rid of the fact that the filamenting region shifts towards shorter distances when the intensity is increased, the results presented for each intensity bin correspond to the region where the filament number was maximum. 


\begin{figure}[ht]
\begin{center}
	\subfigure{\includegraphics[trim={1cm 8.7cm 1cm 11cm},clip,width=0.9\columnwidth]{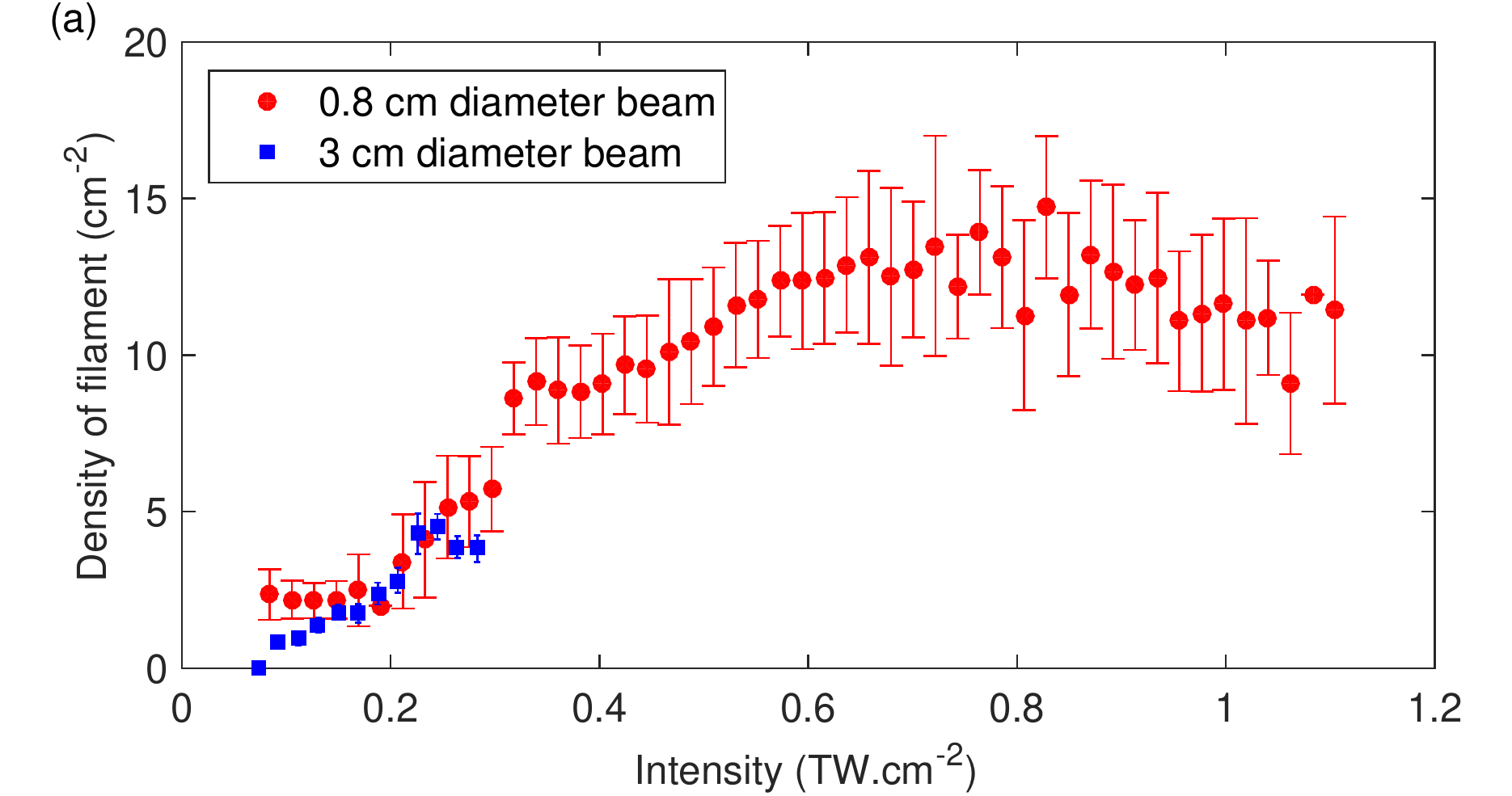}}\\
	\subfigure{\includegraphics[trim={0cm 0cm 0.7cm 0.5cm},clip,width=1\columnwidth]{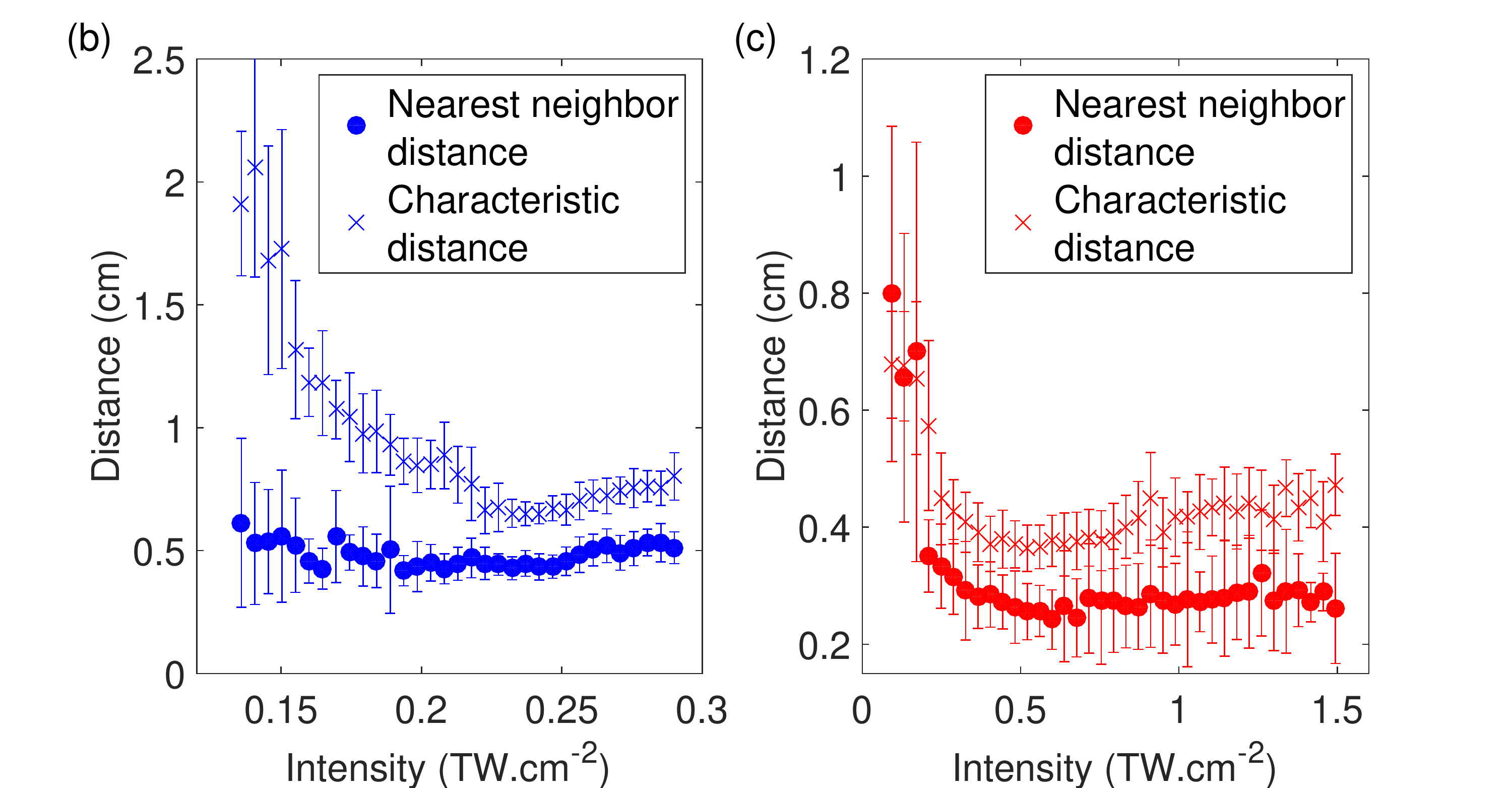}}	
\end{center}
	\caption{Experimental averaged filament density in the beam profile as a function of the averaged intensity (a), for the two beam geometries of our multi-Terwatt near-infrared laser. 
	Mean nearest-neighbor distance and characteristic distance $d={1}/{\sqrt{\sigma}}$, where $\sigma$ is the mean filament density as a function of the incident intensity, in (b) the 3~cm diameter beam , and (c) the 0.8~cm diameter beams.}
	\label{fig:fig1}
\end{figure}

As displayed in Figure \ref{fig:fig1}(a), the beam-averaged filament density rises almost linearly (R$^2$~=~0.94 for all points below 0.4~TW/cm$^2$) with the beam-averaged incident intensity, until it saturates at a typical packing density of 12~cm$^{-2}$ for intensities above $\approx 0.4$~TW/cm$^2$.
These saturation density and sub-TW/cm$^2$ intensity values are in line with what was estimated by~\cite{HeninPKWJKBSSNSRWSMBS2010a}, i.e. 10~cm$^{-2}$ and 0.2~TW/cm$^2$. The fact that the present experiment is performed at a 100 times lower peak power and beam cross section than those of~\cite{HeninPKWJKBSSNSRWSMBS2010a} demonstrates the universality of the saturation of the filament density at high intensities, regardless of the influence of the beam boundaries.

This transition to the regime where the interaction between filaments dominates (solid-like phase) appears to imply a rich physics when comparing our data with previous studies, traditionally expressed in terms of number of critical powers per filament (with $P_\mathrm{cr}$ = 4 GW  \cite{HeninPKWJKBSSNSRWSMBS2010a,ChinPLINKKA2002,StelmRMYSKAWW2004}). 
For intensities exceeding the 0.4~TW/cm$^2$ threshold for condensation, we observe that the number of critical powers per filament increases linearly with the incident intensity due to the saturation of the filament number. 
Conversely, below 0.06~TW/cm$^2$, various multifilamentation data from the literature (\cite{M'ejKYSFWSVNCB2005,SkupiBPLMYKSWRWBS2004,M'ecDATFPMCSS2005,B'ejBEMWALSSKBCGBBBCCHLMMMPPRR2007}) fit to the value of 5~$P_\mathrm{cr}$ per filament, typical of the power-limited regime (gas-like phase)~\cite{HeninPKWJKBSSNSRWSMBS2010a}.
Between these two phases, we identify a new mixed phase state, where both gas and solid co-exist. In this regime, the number of power amounts 10 to 20~$P_\mathrm{cr}$ per filament.


To better understand this mixed-phase, characterized by intensities lower than the phase transition threshold, we calculated the mean nearest-neighbor distance of the filaments for energy bins corresponding to 40 single shot pictures of the beam each. We compared this average nearest-neighbor distance with a characteristic inter-filament distance calculated as $d={1}/{\sqrt{\sigma}}$, where $\sigma$ is the filaments surface density in the beam profile.

\begin{figure*}[ht]
\begin{center}
	\label{fig:image_gros}\includegraphics[trim={0cm 0cm 10.5cm 0cm},clip,width=1.5\columnwidth]{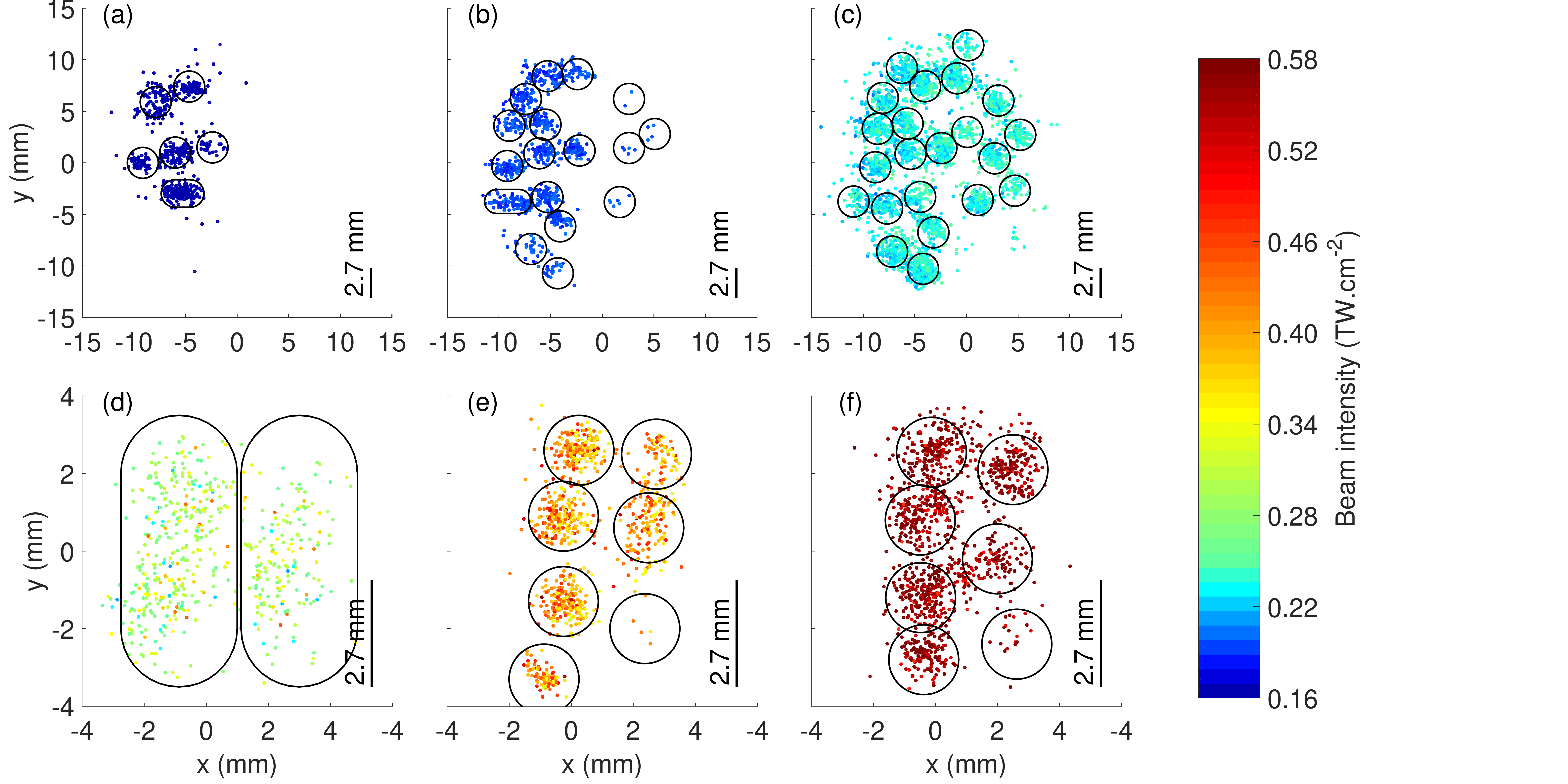}	
\end{center}
	\caption{Superposition of the filaments detected over 200 single-shot images for the 3~cm beam (a-c) and the 8~mm beam (d-f) displayed in three classes of average incident beam intensities for clarity. Each pattern of filaments corresponding to the same incident beam intensity is displayed with the same color. The wandering range region of each individual filament is marked as a black circle. An example of individual pattern is displayed in Figure 1 of the supplementary materials.}
	\label{fig:Images}
\end{figure*}

The transition from the gas phase to the solid phase is evidenced by the decrease of both the nearest-neighbor distance and the characteristic distance $d$ (Figure \ref{fig:fig1}(b-c)). The nearest-neighbor distance decreases to a constant plateau at 2.7~mm when the incident beam intensity is above 0.4~TW/cm$^2$ (Figure \ref{fig:fig1}(c)). 
Below this threshold (Figure \ref{fig:fig1}(b)), the characteristic distance in the 3~cm beam is divided by a factor 3 over the intensity range while the nearest-neighbor distance decreases only by 20\%. 

This means that filaments are not randomly distributed but rather packed into particular regions of the beam. This behavior is illustrated in Figure 2, where each dot corresponds to the occurrence of a single filament over 200 pulses accumulation, and the black circles evidences the shot-to-shot wandering region of each individual filament. Below 0.4~TW/cm$^2$, filaments do not cover the whole beam profile, but form dense clusters with a typical nearest-neighbor distance of 2.7~mm. Increasing intensity mainly leads to the growth of the existing packed regions until the whole beam is saturated. In other words, the local inhomogeneities in the beam profile act as seeds for local nucleation/deposition of the solid phase appearing as packed black circles in Figure \ref{fig:Images}.

In the 8 mm beam, nearest neighbor distance and characteristic distance behave similarly (Figure 1(c)), showing that the mixed phase plays a minor role in the small beam. Furthermore, this phase is harder to evidence due to the low number of filaments (see Figure \ref{fig:Images}(d)). When increasing the intensity, the small size available and the more homogeneous profile allow the formation of a solid pattern covering the whole beam cross-section (Figure \ref{fig:Images}(e-f)).

The nearest-neighbor distance of 2.7~mm corresponds to a density of 13.7~cm$^{-2}$ for a square lattice, and 15.8~cm$^{-2}$ for a close-packed (honeycomb) lattice, slightly above the previously reported beam-averaged space-limited density of 10~cm$^{-2}$~\cite{HeninPKWJKBSSNSRWSMBS2010a}. As the nearest-neighbor distance is representative of the equilibrium distance of the filaments interaction, the fixed value of the former (plateau in Figure \ref{fig:fig1}(c)) indicates that filaments form a solid with a characteristic Wigner-Seitz constant of 2.7~mm. The statistical analysis of the pictures also provides access to the pulse to pulse wandering of each individual filament and thus an analogy of a temperature (linked to the intensity fluctuations in the beam profile). For filaments in the gas phase (i.e. isolated filaments and I~$<$~0.4~TW/cm$^2$) the variance on the position amounts 0.20--0.22~cm$^{-2}$, while for filaments in the solid phase (I~{$>$}~0.4~TW/cm$^2$) the variance drastically decreases to 0.06--0.07~cm$^{-2}$, in agreement with a stronger binding potential between filaments.

The presence of the solid-like interaction between filaments can be modelled by considering the Hamiltonian associated to the two-dimensional nonlinear Schr\"{o}dinger equation, that describes the beam propagation. We consider here its normalised version: 
\begin{equation}
 i \partial_{\eta}\psi + \Delta \psi + f(\vert\psi\vert^2)\psi = 0 
\end{equation}

where $\psi$ is the complex amplitude of the electric field, $\eta$ is the normalised propagation coordinate, $\Delta$ the transverse Laplacian which accounts for geometrical diffraction, and the function $f$ describes the nonlinear mechanisms at play. 

Notice that we focus here on the interaction between already established filaments propagating in a stabilized regime, not their formation during the self-focusing, pre-collapse, phase. We consider in the $f$ function the Kerr effect and the defocusing due to plasma (ionization with K photons):
\begin{equation}
f(\vert\psi\vert^2) = \vert\psi\vert^2 - \vert\psi\vert^{2K}
\end{equation}

The Hamiltonian associated to this equation is $H = \int \mathcal{H} dr $ with:
\begin{equation}
\mathcal{H} = -\frac{\vert \psi \vert ^4}{2} + \vert \nabla \psi \vert ^2 + \frac{\vert \psi \vert ^{2(K+1)}}{K+1} 
\label{eq:hamiltonian}
\end{equation}

Following the pioneering work from L. Berge et al.~\citep{Berge1997}, we then study the interaction between two filaments by considering $\psi$ as the superposition of two localised Gaussian beamlets:

$$ \psi(r,\delta,\theta,N) = \sqrt{\frac{N}{\pi \rho^2}} \left[ e^{-r^2/(2\rho^2) }   + e^{ -(r-\delta) ^2/(2\rho ^2) +i \theta } \right] $$


Where $\delta$ is the distance between the beamlets and $\theta$ their relative phase. For the sake of simplicity, both beamlets have same power $N$ and radius $\rho$. For a filament diameter of 150~$\mu$m, the maximum intensity in the beamlet corresponding to one critical power ($P=P_{cr}$, $N=4\pi$ here \cite{berge_amalgamation_1997}, 4~GW at 800~nm) is 5.6~TW/cm$^2$.
We then calculate numerically the interaction Hamiltonian (using the \textit{chebfun} library in matlab \cite{Townsend2013}):

$$ H_{\text{int}}(\delta,\theta,N) =  H(\delta,\theta,N) - H(\infty,\theta,N) $$

\begin{figure}[h!]
	\centering
		\includegraphics[trim={0cm 0cm 0cm 1cm},clip,width=0.7\columnwidth]{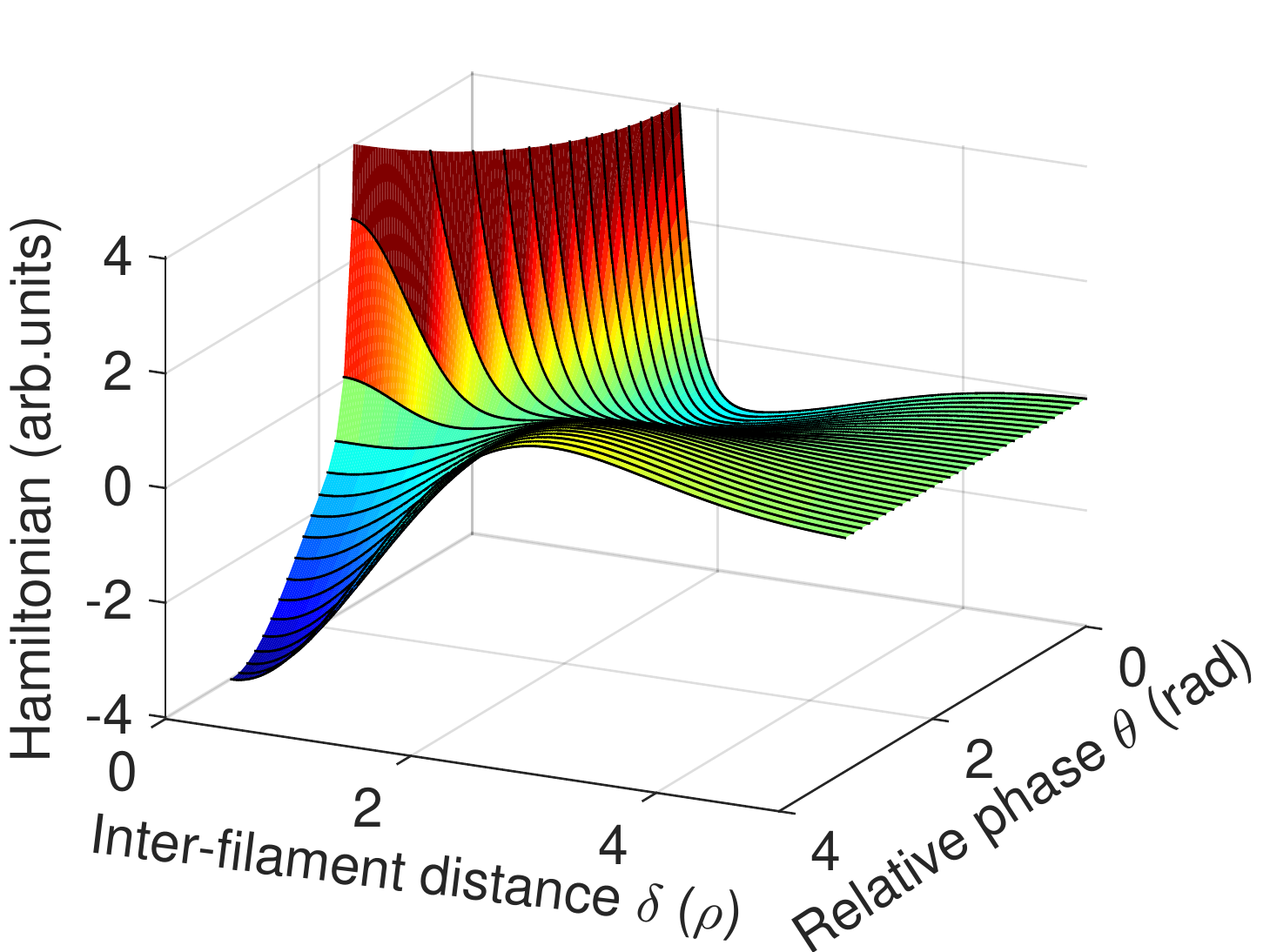} 						
	\caption{Interaction potential between two filaments of power $P=0.2$~$P_{cr}$ as a function of the inter-filament distance $\delta$ and their relative phase $\theta$}
	\label{fig:saddle}
\end{figure}

\begin{figure}[ht]
\begin{center}
\centering
		\includegraphics[trim={0cm 3cm 0cm 1cm},clip,width=1.0\columnwidth]{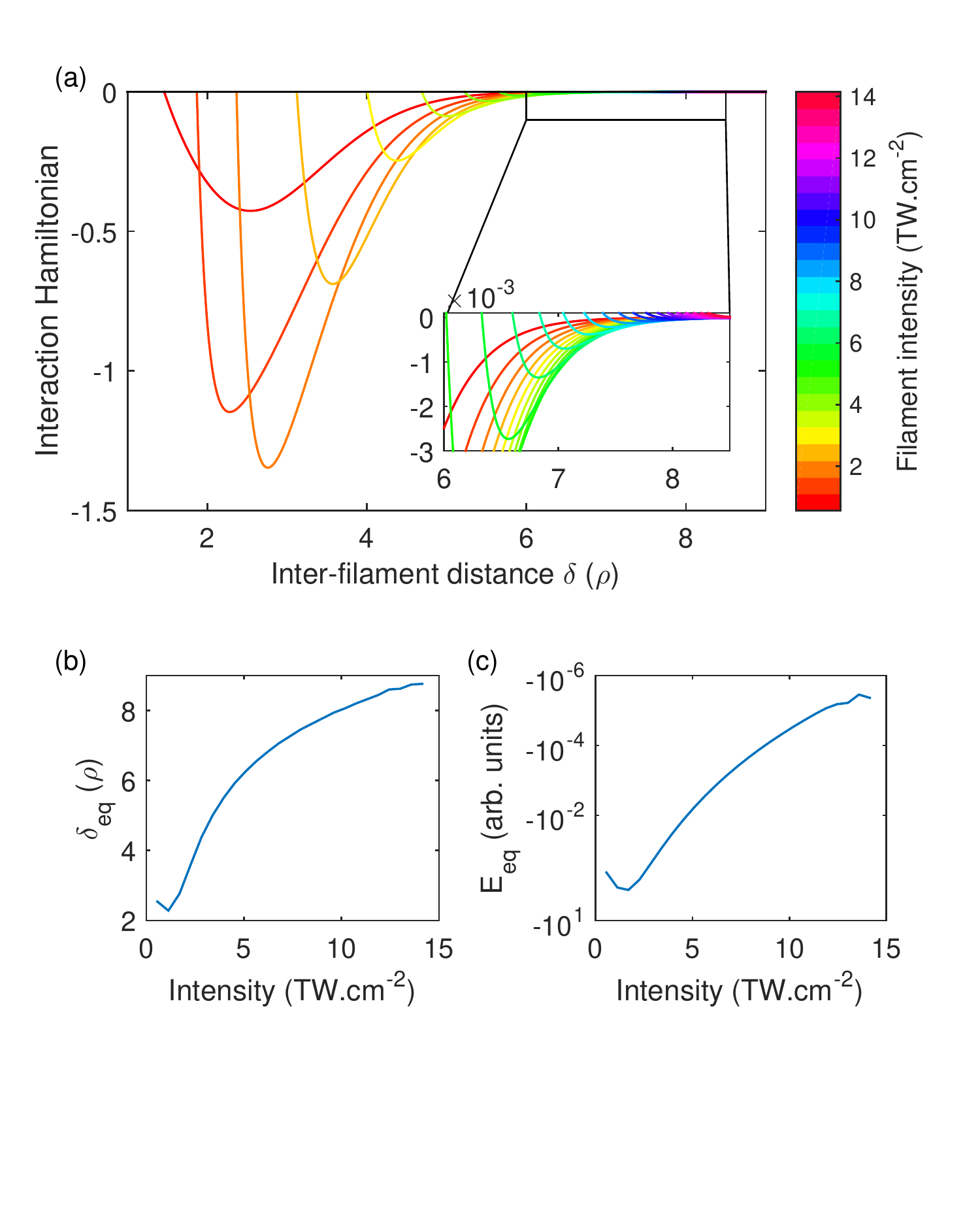}		
\end{center}
	\caption{(a): interaction Hamiltonian for $0<P<2.5 \, P_{cr}$ and $\theta = 0$ function of the inter-filament distance $\delta$. Evolution of (b): the equilibrium distance and (c): the equilibrium Hamiltonian value of the curves displayed in (a).}
	\label{fig:hamilton}
\end{figure}

Figure \ref{fig:saddle} displays the calculated Hamiltonian as a function of both $\delta$ and $\theta$, for a moderate power $P=0.2P_{cr}$ (corresponding to 1.12~TW/cm$^2$ maximum intensity). The curve exhibits a saddle shape, with a turning point for $\theta = \pi/2$ and $\delta \simeq 1.5 \rho$. One side of the saddle ($\theta > \pi/2$) leads to merging of the two filaments (i.e. $H_{\text{int}}$ is min when $\delta = 0$), while the other part ($\theta < \pi/2$) leads to a "Morse-like" potential with an equilibrium distance $\delta_{\text{eq}} > 0$.  
The analogy with a diatomic molecule is quite remarkable. The overall shape of this curve is kept when increasing the power although the amplitudes of the negative and positive peaks for $\delta = 0$ become orders of magnitude larger than the peaks for $\delta > 0$.
The bonding potential observed for $\theta< \pi/2$ is the result of the attractive part due to the Kerr effect (filaments in phase tend to merge because of self-focusing) and the repulsive effect of the plasma (reducing the inter-filament distance creates more plasma that repel the filaments away from each other).
If we now consider that formed and coexisting filaments are in phase~\cite{jhajj_spatiotemporal_2016}, we see in Figure \ref{fig:hamilton}(b) and (c) that $\delta_{\text{eq}}$ increases with the power and tends to converge to $\delta = 9\rho$, that is 1.35~mm for $\rho = 150 \; \upmu$m. This value, although a factor 2 smaller than the observed one (2.7~mm) is still comparable, especially considering that we calculated the interaction between only two filaments (like the bond length in a diatomic molecule) and not a 2D lattice.  Clearly, the equilibrium distance will increase in this latter case (a factor 1.5 is typical between the equilibrium distance in a dimer and the lattice constant in the equivalent solid).
The energy gained by forming a solid-like structure reduces when the power increases (Figure \ref{fig:hamilton}(a)), because of the dominant repulsive effect of the plasma.

Although the values of the transition thresholds and of the inter-filaments distances are specific to NIR pulses propagating at atmospheric pressure, the phenomenon itself (transition between different phases of filamentation) and its description are valid for a wide range of mutlifilamentation processes in the sense that it requires only the competition of self-focusing and self-defocusing effects.

It should be noted that although the present model is able to qualitatively explain the essence of the observed phenomenons, modelling filaments with a gaussian shape and a uniform phase is a crude model. Further modelling should include realistic intensity flux \cite{faccio_conical_2006,faccio_experimental_2009} and intensity shapes \cite{kolesik_dynamic_2004}, and advanced phase profile such as the recently studied spatio-temporal optical vortices \cite{jhajj_spatiotemporal_2016}. Such features will expectedly improve the description of inter-filament interactions.

We have investigated the transition between a multifilamentation regime with weak interfilament interaction (gas-like) to a regime with strong interfilament  interaction (solid-like). This transition is driven by the incident local beam intensity, which plays the role of pressure in our analogy. We evidence a mixed phase state occurring for intensities between 0.06~TW/cm$^2$ and 0.4~TW/cm$^2$ for a near-infrared beam, 
in which filaments are closely packed in localized clusters, probably nucleated on inhomogeneities in the transverse intensity profile of the beam. The density in these clusters is almost constant, while the remaining of the beam, initially void of filaments, progressively fills up as the clusters spread over the whole profile when the incident intensity is increased.

These findings are confirmed by considering and calculating the interaction Hamiltonian between two filaments, taking into account the effect of diffraction, Kerr and plasma(see Equation \ref{eq:hamiltonian}). This simple model suggests that filaments interact with an effective potential, which shares fundamental characteristics with an atomic model: the shape of the potential is the result of the counteracting actions of the short-range repulsion of the plasma (defocusing) and the longer-range attraction of the Kerr effect (focusing). The resulting bond length (equilibrium distance) is a signature of this trade-off. Similarly, the potential curve is dependant on the relative phase between the two interacting filaments. This behaviour reminds a diatomic molecular orbital (MO) formed by linearly superposing the two orbitals of the individual atoms (AO) in the orbital approximation: the relative phase between the two AO determines indeed whether the molecular state will be bonding or anti-bonding.

\begin{acknowledgements}

The authors aknowledge support from the European Research Council Advanced Grant ``Filatmo''.
The technical help of M. Moret was highly appreciated. Authors would like to thank D. Eeltink for the help during experiment, and W. Ettoumi for the fruitful discussions.
\end{acknowledgements}

\end{document}